\documentstyle[twoside]{article}

\catcode`\@=11
\long\def\@makefntext#1{ 
\protect\noindent \hbox to 3.2pt {\hskip-.9pt
$^{{\eightrm\@thefnmark}}$\hfil}#1\hfill} 

\def\thefootnote{\fnsymbol{footnote}}
 \def\@makefnmark{\hbox to 0pt{$^{\@thefnmark}$\hss}}  

\def\ps@myheadings{\let\@mkboth\@gobbletwo
\def\@oddhead{\hbox{} 
\rightmark\hfil\eightrm\thepage}
\def\@oddfoot{}\def\@evenhead{\eightrm\thepage\hfil 
\leftmark\hbox{}}\def\@evenfoot{}
\def\sectionmark##1{}\def\subsectionmark##1{}}

\oddsidemargin=\evensidemargin
\renewcommand{\thefootnote}{\fnsymbol{footnote}}

\newcounter{sectionc}\newcounter{subsectionc}\newcounter{subsubsectionc}
\renewcommand{\section}[1] {\vspace{12pt}\addtocounter{sectionc}{1}
\setcounter{subsectionc}{0}\setcounter{subsubsectionc}{0}\noindent
        {\tenbf\thesectionc. #1}\par\vspace{5pt}}
\renewcommand{\subsection}[1] {\vspace{12pt}\addtocounter{subsectionc}{1}
        \setcounter{subsubsectionc}{0}\noindent
        {\bf\thesectionc.\thesubsectionc. {\kern1pt \bfit #1}}\par\vspace{5pt}}
\renewcommand{\subsubsection}[1]
{\vspace{12pt}\addtocounter{subsubsectionc}{1}
        \noindent{\tenrm\thesectionc.\thesubsectionc.\thesubsubsectionc.
        {\kern1pt \tenit #1}}\par\vspace{5pt}}
\newcommand{\nonumsection}[1] {\vspace{12pt}\noindent{\tenbf #1}
        \par\vspace{5pt}}

\newcounter{appendixc}
\newcounter{subappendixc}[appendixc]
\newcounter{subsubappendixc}[subappendixc]
\renewcommand{\thesubappendixc}{\Alph{appendixc}.\arabic{subappendixc}}
\renewcommand{\thesubsubappendixc}
        {\Alph{appendixc}.\arabic{subappendixc}.\arabic{subsubappendixc}}

\renewcommand{\appendix}[1] {\vspace{12pt}
        \refstepcounter{appendixc}
        \setcounter{figure}{0}
        \setcounter{table}{0}
        \setcounter{lemma}{0}
        \setcounter{theorem}{0}
        \setcounter{corollary}{0}
        \setcounter{definition}{0}
        \setcounter{equation}{0}
        \renewcommand{\thefigure}{\Alph{appendixc}.\arabic{figure}}
        \renewcommand{\thetable}{\Alph{appendixc}.\arabic{table}}
        \renewcommand{\theappendixc}{\Alph{appendixc}}
        \renewcommand{\thelemma}{\Alph{appendixc}.\arabic{lemma}}
        \renewcommand{\thetheorem}{\Alph{appendixc}.\arabic{theorem}}
        \renewcommand{\thedefinition}{\Alph{appendixc}.\arabic{definition}}
        \renewcommand{\thecorollary}{\Alph{appendixc}.\arabic{corollary}}
        \renewcommand{\theequation}{\Alph{appendixc}.\arabic{equation}}
        \noindent{\tenbf Appendix \theappendixc #1}\par\vspace{5pt}}
\newcommand{\subappendix}[1] {\vspace{12pt}
        \refstepcounter{subappendixc}
        \noindent{\bf Appendix \thesubappendixc. {\kern1pt \bfit #1}}
        \par\vspace{5pt}}
\newcommand{\subsubappendix}[1] {\vspace{12pt}
        \refstepcounter{subsubappendixc}
        \noindent{\rm Appendix \thesubsubappendixc. {\kern1pt \tenit #1}}
        \par\vspace{5pt}}

\newcommand{\textlineskip}{\baselineskip=13pt}
\newcommand{\smalllineskip}{\baselineskip=10pt}

\def\eightcirc{
\begin{picture}(0,0)
\put(4.4,1.8){\circle{6.5}}
\end{picture}}
\def\eightcopyright{\eightcirc\kern2.7pt\hbox{\eightrm c}}

\newcommand{\copyrightheading}[1]
        {\vspace*{-2.5cm}\smalllineskip{\flushleft
        {\eightrm International Journal of Modern Physics D, #1}\\
        {\eightrm $\eightcopyright$\, World Scientific Publishing
         Company}\\
         }}

\newcommand{\publisher}[2]{{\begin{center}\eightrm\smalllineskip
        Received #1
        \end{center}
        }}

\def\abstracts#1#2#3{{
        \centering{\begin{minipage}{4.5in}\baselineskip=10pt\eightrm
        \centerline{ABSTRACT}
        \parindent=0pt #1\par
        \parindent=15pt #2\par
        \parindent=15pt #3
        \end{minipage} }\par}}

\newcommand{\bibit}{\nineit}

\renewenvironment{thebibliography}[1]                   
        {\ninerm
         \baselineskip=11pt                             
         \begin{list}{\arabic{enumi}.}
        {\usecounter{enumi}\setlength{\parsep}{0pt}
         \setlength{\leftmargin 17pt}{\rightmargin 0pt} 
         \setlength{\itemsep}{0pt} \settowidth          
        {\labelwidth}{#1.}\sloppy}}{\end{list}}

\newcounter{itemlistc}
\newcounter{romanlistc}
\newcounter{alphlistc}
\newcounter{arabiclistc}

\newcommand{\fcaption}[1]{
        \refstepcounter{figure}
        \setbox\@tempboxa = \hbox{\eightrm Fig.~\thefigure. #1}
        \ifdim \wd\@tempboxa > 5in
           {\begin{center}
        \parbox{5in}{\eightrm \smalllineskip Fig.~\thefigure. #1 }
            \end{center}}
        \else
             {\begin{center}
             {\eightrm Fig.~\thefigure. #1}
              \end{center}}
        \fi}

\newcommand{\tcaption}[1]{
        \refstepcounter{table}
        \setbox\@tempboxa = \hbox{\eightrm Table~\thetable. #1}
        \ifdim \wd\@tempboxa > 5in
           {\begin{center}
        \parbox{5in}{\eightrm\smalllineskip Table~\thetable. #1 }
            \end{center}}
        \else
             {\begin{center}
             {\eightrm Table~\thetable. #1}
              \end{center}}
        \fi}

\def\@citex[#1]#2{\if@filesw\immediate\write\@auxout    
        {\string\citation{#2}}\fi                       
\def\@citea{}\@cite{\@for\@citeb:=#2\do                 
        {\@citea\def\@citea{,}\@ifundefined             
        {b@\@citeb}{{\bf ?}\@warning
        {Citation `\@citeb' on page \thepage \space undefined}}
        {\csname b@\@citeb\endcsname}}}{#1}}

\newif\if@cghi
\def\cite{\@cghitrue\@ifnextchar [{\@tempswatrue
        \@citex}{\@tempswafalse\@citex[]}}
\def\citelow{\@cghifalse\@ifnextchar [{\@tempswatrue
        \@citex}{\@tempswafalse\@citex[]}}
\def\@cite#1#2{{$\null^{#1}$\if@tempswa\typeout
        {IJCGA warning: optional citation argument
        ignored: `#2'} \fi}}

\def\pmb#1{\setbox0=\hbox{#1}
        \kern-.025em\copy0\kern-\wd0
        \kern.05em\copy0\kern-\wd0
        \kern-.025em\raise.0433em\box0}

\def\fnt#1#2{\footnotetext{\kern-.3em
        {$^{\mbox{\scriptsize #1}}$}{#2}}}

\def\fpage#1{\begingroup
\voffset=.3in
\thispagestyle{empty}\begin{table}[b]\centerline{\footnotesize #1}
        \end{table}\endgroup}

\def\runninghead#1#2{\pagestyle{myheadings}
\markboth{{\eightit{\quad #1}}\hfill}{\hfill{\eightit{#2\quad}}}}

\font\tenbf=cmbx10
\font\tenit=cmti10
\font\tenit=cmti10
\font\bfit=cmbxti10 at 10pt
 1
 1
 1

\font\ninerm=cmr9
\font\nineit=cmti9

\font\eightrm=cmr8
\font\eightit=cmti8

\def\qed{\hbox{${\vcenter{\vbox{                          
   \hrule height 0.4pt\hbox{\vrule width 0.4pt height 6pt
   \kern5pt\vrule width 0.4pt}\hrule height 0.4pt}}}$}}

\runninghead{Jerzy Lewandowski} {Topological Measure
on the Space of Connections $\ldots$}

\def\A{{\cal A}}
\def\G{{\cal G}} \def\C{{\cal C}} \def\Cb{{\bar {\cal C}}}
\def\AG{\A/\G} \def\AGb{\overline{\AG}}
\def\S{\Sigma}
\def\L{{\cal L}}
\def\g{\gamma} \def\a{\alpha} \def\b{\beta}
\def\gt{{\tilde \g}} \def\ft{{\tilde f}} \def\Pt{{\tilde P}}
\def\Tr{{\rm Tr}}
\def\l{\lambda}

\begin{document}
\normalsize\textlineskip
{\thispagestyle{empty}
\setcounter{page}{1}

\renewcommand{\thefootnote}{\fnsymbol{footnote}} 

\copyrightheading{Vol. 0, No. 0 (1993) 000--000}

\vspace*{0.88truein}

\fpage{1}
\centerline{\bf TOPOLOGICAL MEASURE AND GRAPH-DIFFERENTIAL GEOMETRY}
\vspace*{0.035truein}
\centerline{\bf ON THE QUOTIENT SPACE OF CONNECTIONS
\footnote{This work was
supported in part by the  National Science Foundation grant
PHY91-07007 and
Polish Committee for Scientific Research (KBN) through grant no.
2 0430 9101.}}
\vspace{0.37truein}
\centerline{\eightrm JERZY LEWANDOWSKI\footnote{Permanent address: Instytut
 Fizyki Teoretycznej, Uniwersytet Warszawski,
ul. Hoza 69, 00-689 Warszawa, Poland}}
\vspace*{0.015truein}
\centerline{\footnotesize\it Department of Physics, University of Florida}

\baselineskip=10pt
\centerline{\footnotesize\it  Gainesville, FL 32611, USA}
\vspace{0.225truein}
\publisher{September 1, 1993}

\vspace*{0.21truein}
\abstracts{\noindent Integral calculus on the space $\AG$ of
gauge equivalent connections is developed. By carring out a non-linear
generalization  of the theory of cylindrical measures on topological
vector spaces, a faithfull, diffeomorphism invariant measure
is introduced on a suitable completion of $\AG$. The strip
(i.e. momentum) operators are densely-defined in the resulting
Hilbert space and interact with the measure correctly}{}{}

\vspace*{-3pt}\textlineskip
\section{Introduction}
\noindent

The space ${\cal A}/{\cal G}$ of gauge equivalent connections plays a
central role in gauge theories as well as in the connection-dynamics
formulation of general relativity$^1$. The problem of constructing quantum
kinematics of these theories can therefore be reduced to: i)Introducing
the algebra of a complete set of manifestly gauge invariant observables;
and, ii) Finding a suitable representation of this algebra. For the first
step, one can use the loop-strip variables introduced by Rovelli, Smolin
and Ashtekar$^2$. A general strategy to complete the second step was developed
by Ashtekar and Isham$^3$ using the Gel'fand theory of representations of $C^*$
algebras. We are able to carry this strategy to completion by introducing
a {\it diffeomorphism invariant} measure on (an appropriate completion of)
${\cal A}/{\cal G}$, using the resulting space of square-integrable functions
on this space as quantum states and introducing techniques from differential
geometry on this space to represent the strip operators. These results
provide a rigorous kinematical framework for the quantum version of these
theories.

Graphs, loops, connections and holonomies are used as the main tools.
The measure is introduced via a non-linear generalization of the theory
of cylindrical functions on topological vector spaces. In addition to being
diffeomorphism invariant, to our knowledge, it is the first faithful
measure on (the completion of) ${\cal A}/{\cal G}$. Similar techniques enable
us to do differential geometry on this space in terms of a family of
projections onto finite dimensional manifolds. The resulting family of strip
operators (representing momenta) are shown to be symmetric with respect
to this measure. This framework, in particular, enables one to define
a Rovelli-Smolin$^4$ loop transform in a rigorous fashion, thereby leading to
a loop representation for quantum states. Furthermore, the loop (i.e.
holonomy) and strip (i.e. conjugate momentum) operators are well defined
also in the loop representation. Finally, the framework offers
a new and rigorous avenue to regularization of several physically
interesting operators. Mathematically, it opens a new avenue to a
graph-theory of measures, differential geometry on the space of (gauge
equivalent) connections, and a relation between knot and link invariants
and measures on this space.

We consider here the space $\A$ of $G$-connections defined over a
 manifold $\S$. $\AG$ is the quotient with respect to the group $\G$
 of gauge transformations on $\A$. In application to gravity, $\S$ is
 a Cauchy 3-surface and a gauge group $G$ is taken to be $SU(2)$.
Nonetheless, most of our results   remains true if we assume only that
 $G$ is a compact Lie group.

Below we give an outline of our technics.
{\bibit This research was carried out in colaboration with
Abhay Ashtekar. Details will appear in a joined paper.}$^5$.

\section{Graph-manifold structure on $\AG$.}
\noindent
We introduce on $\AG$ a family of projections $\pi:\AG\ \
\longrightarrow G^n/{\rm ad}$
where $n$ takes all the natural values depending on $\pi$
and an element of $G^n/{\rm ad}$ is the class of conjugated elements
of $G^n$ ( $(g_1,...,g_n) \sim (a^{-1}g_1a,...,a^{-1}g_n a)$).
The projections which we construct are labelled by {\it trivialised
graphs}. We fix in $\S$ a reference point $x_0$ and consider the set
$\L$ of all the piecewise analytic loops which begin at $x_0$.
Let $\g$ be a graph analytically embedded into $\S$. It consists
of two sets:  $E$ of edges and $V$ of vertexes. Connect each vertex
$v$ with $ x_0$ by an analytic path $q(v)$ which does not overlap any
of the edges. A trivialization\ $\gt$ of a graph $\g$ is the
map
$$
E\ni e\ \ \mapsto\ \ \b(e) \ \ =\ \
 q(v_+(e))^{-1}
\circ e\circ
q(v_-(e))\in \L,
$$
where the vertex $v_-(e)$ and $v_+(e)$ is the beginning and the end of
$e$ respectively. A trivialized graph $\gt$ defines the projection
$$
\AG\ni A \ \ {\buildrel \pi_\gt \over \mapsto}\ \
[(H(\b(e_1),A),
..., H(\b(e_n),A))]\in G^n/{\rm ad}\eqno(1),$$

with $H(\b,A)$ denoting the parallel transport with respect to
 a connection $A$ around a loop $\b$ ($n$ is the number of edges
 in $\g$).

Below we present two examples of geometrical objects
of our theory:
cylindrical functions and a topological measure.

Given a map $\pi_\gt$  (1) we can lift any continuous
function $\ft$ defined on $G^n/{\rm ad}$ to a function
$f = \pi_\gt^*\ft$ defined on $\AG$. A function $f$
which can be obtained in this way by using an analytically
embedded graph will be called
{\it cylindrical function}. We consider only such functions
$\ft$ for which $f$ does not depend on trivialization
of $\g$. Denote by $\C$ the set of cylindrical functions.
First important property of $\C$ is   that {\it if $f,g\in \C$
then both $f+g,\ fg\in \C$ $^5$}. Hence $\C$ is an algebra.
We can equip it with a norm $\|f\|:= sup |f|$. A principal example
of a cylindrical function is a "Wilson loop", the function defined
by any loop $\a$ in $\S$:
$$T_\a (A) \ =\ \Tr H(\a,A).$$
The second important result$^5$ is that "Wilson loops" span a vector
 space dense in $\C$. Therefore, the completion of $\C$,
 $(\Cb, \|\ \|, *)$ (the *-operation being a complex conjugation)
 is isomorphic to the Ashtekar-Isham $C^*$
holonomies algebra}.

Our equivalent construction of the A-I algebra provides us
automatically with a natural functional defined on $\AG$. Indeed,
 it turns out$^5$   that
for every cylindrical function $f$ the quantity
$$\int_{G^n}\ft d^n g\ =:\ \int_{\AG} f\eqno(2)
$$
{\it is independent of which projection compatible with
$f$ we use},
where $\ft$ was naturally extended from $G^n/{\rm ad}$
to $G^n$ and $dg$ is the normalized Haar measure on
$G$. The functional $\int_{\AG}$, as the notation suggests,
may be thought of as a generalized measure on $\AG$. On the
 other hand, since  $\int_{\AG}$ {\it is strictly positive
 and continuous$^5$ on $\C$} it defines a regular and faithful
 measure $d\mu$ on the A-I   extension $\AGb$ for $\AG$, the spectrum
of the algebra. In addition, for our definition of the measure
 we did not use
any extra structure on $\S$. This is reflected in the topological
character of the measure: $\int_{\AG}$ is invariant with respect
to the diffeomorphisms acting on $\S$.
In this way the Hilbert space $L^2(\AGb, d\mu)$ of quantum states
for a topological theory on $\AG$ has been constructed.

It is worth noting, that our measure, although arises naturally from the
Haar measure on $G$, is not the only measure which fits  the projections
(1). Recently, Baez  has exploited this general strategy to introduce
other, more involving the graph theory,  examples of  graph measures.$^6$

\section{The momentum operator}
\noindent
Unlike the measure, a strip operator which represents momentum
was introduced before$^2$  directly on $\AG$. Our graph-differential
geometry technics however are successfully applied for an evaluation of the
Hermitian conjugation of this operator.

To begin with, consider a momentum variable $E_i^a(x)$ canonically
conjugate to the connection variable $A_a^i(y)$ ($a$ is a space index
and $i$ is a Lie algebra index). In gravity $E^a_i$ is a sensitized
tried. Equivalently, we represent $E^a_i$ by a matrix valued 2-form $E$
such that
$$\int_\S \Tr A\wedge E \ =\ \int_\S d^3 xA_a^iE_i^a.$$
A strip $S$ means an analytical embedding
$S^1\times [0,1]\ni(s,\l)\ \mapsto\  S(s,\l)\in\S$. To every point
$x=S(s,\l)$ we assign the $G$ valued function of $A$,
$H(x,A)\ =\ H(\b_{x}, A)$
where $\b_{x}(t):=S(t+s,\l)$ is the loop passing through
$x$. The {\it strip variable} is a gauge invariant integral of $E$ on $S$,
$$P_S(A,E)\ =\ \int_S \Tr(H(x,A)E(x)).$$
The strip operator corresponding to $P_S$
is the "distributional vector field" defined on $\AG$ by
the usual  Poisson bracket:
$$P_S(f(A))\ =\ \{P_S(A,E),f(A)\}$$
What is important for us, is that $P_S$ carries every $T_\a$ function
into a linear superposition of other $T_\b$s. Hence, from our point of view,
the operator is  is densely defined in  $L^2(\AGb, d\mu)$.

Now, we are coming back to the graph-differential geometry approach.
A graph $\g$ is {\it compatible
with a strip} $S$ if the following two conditions are satisfied ({\it i})
 every  edge either intersects $S$
at most at its ends or is entirely contained in $S$, ({\it ii}) if a vertex
$v$ belongs to $S$ then the loop $\b_v$ is contained in the
graph $\g$.
Here are our results: ({\it a})  Suppose  $\g$ is compatible with a strip $S$;
then there exists on the corresponding $G^n$ a vector field
$\Pt_S$ such that for every cylindrical function $f$ compatible
with $\g$ (and such that $\ft$ is differentiable)
$$P_S(f) \ =\ \pi_\g^*\Pt_S(\ft)$$
({\it b}) for every strip operator and a cylindrical function
there exists a graph simultaneously compatible with both of them.

We shall not write here the full expression for $\Pt_{S}$ but we
summarize below its  properties which are relevant.
The component $\Pt_{S(e)}$ on the copy of $G$ corresponding to
an edge $e$ is zero if $e$ does not intersect $S$; otherwise
$\Pt_{S(e)}$ is a conformal Killing vector in $G$.
The total divergence of $\Pt_{S}$  vanishes.  Finally, the map
$P_S\ \mapsto \ \Pt_S$ preserves the Hermitian conjugations
and extends naturally to the commutators. Therefore, since
${\rm div}\Pt_S=0$,
$$P_S^+\ =\ \pi_\g^*(\Pt_S^+)\ = \ -P_S.$$
This  extends to the full Lie algebra generated by all the strip
operators.

\nonumsection{References}

 \end{document}